# Quantum critical scaling for finite temperature Mott-like metal-insulator crossover in a few layered-MoS$_2$

Byoung Hee Moon[1,2†]*, Gang Hee Han[1,3†], Miloš M. Radonjić[4], Hyunjin Ji[2], Vladimir Dobrosavljević[5]*

[1]Center for Integrated Nanostructure Physics, Institute for Basic Science (IBS), Suwon 16419, Republic of Korea

[2]Department of Energy Science, Sungkyunkwan University, Suwon 16419, Republic of Korea

[3]Department of Physics, Incheon National University, Incheon 22012, Republic of Korea

[4]Scientific Computing Laboratory, Center for the Study of Complex Systems, Institute of Physics Belgrade, University of Belgrade, Pregrevica 118, 11080 Belgrade, Serbia

[5]Department of Physics and National High Magnetic Field Laboratory, Florida State University, Tallahassee, Florida 32306, USA

*e-mail: ibhmoon@skku.edu; vlad@magnet.fsu.edu

**The possibility of the strong electron-electron interaction driven insulating phase from the metallic phase in two-dimensions has been suggested for clean systems without intentional disorder, but its rigorous demonstration is still lacking. Here, we examine the finite-temperature transport behavior of a few layered-MoS$_2$ material in the vicinity of the density-driven metal-insulator transition (MIT), revealing previously overlooked universal features characteristic of strongly correlated electron systems. Our scaling analysis, based on the Wigner-Mott theoretical viewpoint, conclusively demonstrates that the transition is driven by strong electron-electron interactions and not disorder, in striking resemblance to what is seen in other Mott systems. Our results provide compelling evidence that transition-metal dichalcogenides provide an ideal testing ground for the study of strong correlation physics, which should open an exciting avenue for future research, making a parallel with recent advances in twisted bilayer graphene.**

Since the first experimental demonstrations[1-5] of a metal-insulator transition in two-dimensional systems (2DMIT) in Silicon, the pivotal role of electron-electron interactions has been recognized[5] as a key factor. By demonstrating that the salient features of this transition are most pronounced in the cleanest samples, more recent experimental studies[6] provided further evidence that this phenomenon is dominated by this strong electronic correlations - and not disorder, in contrast to early theoretical ideas[7,8]. Despite these advances, several basic issues continued to cause controversy, and the progress remained slow. Most importantly, it is still not completely clear if the basic physical mechanism driving this transition bears much in common with other examples of correlated quantum matter, or is it simply an exotic phenomenon specific to two-dimensional electron gases (2DEG) in semiconductors. To place 2DMIT in proper perspective, and to more deeply understand the processes underpinning its striking phenomenology, it is of absolute importance to seek other materials and systems displaying similar behavior. To do this, one needs to establish precisely which aspects of 2DMIT represent the universal features of correlation-driven metal-insulator transitions, and which ones are specific to a given material-dependent realization.

A physical picture suggesting how strong electronic correlations could drive such a transition is based on the mechanism for Wigner crystallization, leading to the so-called Wigner-Mott scenario for 2D-MIT[9-14]. According to this view, in high mobility samples of 2D diluted electron gasses disorder does not play a dominant role for electron localization, except to perhaps help stabilize (pin) the Wigner crystal which is expected to be formed at the lowest densities. The Wigner crystal can be regarded as Mott insulator where carriers form spin 1/2 bound states in the potential well created by Coulomb repulsion of surrounding carriers, so called Wigner-Mott insulator. At higher densities the increased kinetic energy of carriers is eventually able to overcome the Coulomb repulsion, and the Wigner-Mott insulator gives way to the formation of a strongly correlated electron liquid. This perspective, while appealing, has remained controversial, because a controlled microscopic theory is not yet available for the (continuum) 2DEG, in contrast to more conventional lattice models, where more reliable theoretical results have already started to emerge in the last ten years.

Some further progress in this general direction was achieved by recent studies on 2D molecular organic materials at half-filling (so-called Mott organics)[15], where systematic studies of the correlation-driven metal-insulator transitions were performed, providing intriguing new information. These systems are Mott insulators under ambient conditions, but display a transition to a metallic state under moderate pressures, allowing precise and systematic studies of the Mott transition. Most importantly, they are well described by the simplest model of electronic correlation - the single band Hubbard model, making possible direct and very successful comparison with a theory. Here transport experiments revealed several remarkable features, including the scaling of resistivity curves around the so-called Quantum Widom Line, as well as the characteristic evolution of the resistivity maxima, confirming in spectacular detail the predictions of recent theoretical studies[16]. Some similarity of this phenomenology with that of 2DMIT in other 2DEG systems has been noted[14], but the precise connection remained controversial and incomplete.

A remarkable new opportunity to shed light on the fundamental mechanism for 2DMIT was achieved by the recent discovery of a novel class of two-dimensional materials based on transition-metal elements, which already displayed a number of spectacular features[17,18]. These materials allow for unique control over material properties, due to their weak van der Waals bonding between monolayers, allowing easy exfoliation and unprecedented device fabrication. In this paper, we focus on a careful and precise study of transport in a two-dimensional electron gas in a few-layer $MoS_2$ material, and we analyze the results in the light of the Mott-transition scenario guided by theory, as inspired by the recent successes in Mott organics[15]. The results, which we present below, display remarkable similarity to other 2DEG systems, but our analysis demonstrates that its universal features are, in fact, identical to those established for 2D Mott organics. This firmly establishes the strong electronic correlations as the dominant driving force for the metal-insulator transition, painting a remarkably elegant yet robust physical picture concerning its fundamental character. In the following, we first briefly review what is known for the already established for the known examples of bandwidth driven Mott transitions, and then present our experimental data and carry out the appropriate analysis guided by Mott transition theory.

Concerning the phenomenology of the Mott transition, much controversy has existed ever since the early days of high-$T_c$ superconductivity, but more recent experimental and theoretical work established certain robust features, as follows. First, the low-temperature transition has first-order character and displays a phase coexistence region below the critical-end-point at $T = T_c$ so, strictly speaking, it does not display a quantum critical point. However, since $T = T_c$ is an extremely small temperature scale, of the order of only a few percent of the Fermi temperature, the behavior at $T \gg T_c$ features a very well developed Quantum Critical (QC) region displaying universal scaling behavior of the resistivity curves. This QC region is centered around the so-called Quantum Widom Line (QWL), which shows characteristic back-bending at higher temperatures. Experimentally, the QWL is identified by locating the inflection point of the resistivity curves obtained by tuning through a transition at fixed temperature. This quantum critical scaling behavior, which was theoretically first predicted in a seminal 2011 paper[16], was soon confirmed in spectacular detail in a series of experiments on Mott organics by Kanoda and collaborators[15,19], including the characteristic "mirror symmetry" of the scaling function, as also predicted by theory.

Further out on the metallic side, the family of curves were predicted to display resistivity maxima at $T = T_c \times \delta$, marking the onset of quasiparticle transport and the formation of the strongly correlated electron liquid at lower temperatures. Here, the parameter $\delta$ measures the distance from the MIT, which can be accessed by tuning the bandwidth (pressure) or the carrier density in the system. These theoretical predictions provided a clear and precise guidance for experiments, defining a procedure that should be utilized to detect the effects of strong correlations in the vicinity of the Mott transition[14]. The experimental work has largely confirmed essentially all the expected features of the MIT critical region, in those systems where the MIT undeniably has Mott character[20]. The remaining challenge, therefore, is to seek further validation of this scenario in other systems where Mottness is less obvious as the controlling mechanism, which is what we achieve in the present study.

**Experimental data and analysis.**

We now explore the possible presence of Mott quantum criticality in a few layered molybdenum disulfide (MoS$_2$) following a proposed procedure based on theory. In general, 2D materials require the substrate, which strongly restricts the mobility. On the other hand, the ineffective screening enhances interactions between electrons. Furthermore, the large effective electron mass, $m^* \sim 0.5 m_e$, and the small dielectric constant $\varepsilon \sim 8$ for multilayer[21], giving the large dimensionless interaction parameter $r_s \equiv E_C/E_F \propto m^*/\varepsilon\sqrt{n_{2D}} \sim 9$ at $n_{2D} \sim 3 \times 10^{12}\,\text{cm}^{-2}$, are the favorable factors for the observation of Mott-like transition in MoS$_2$. Here, $m_e$ is the free electron mass, $E_C$ is the Coulomb energy, $E_F$ is the kinetic (Fermi) energy, and $n_{2D}$ is the carrier density. In this n-type 2D semiconducting material, the Mott-like transition is pictured as the one from the strongly correlated metallic electron liquid state to the charge ordered insulating state (Wigner crystallization) as the carrier density decreases in the conduction band, as shown in Fig. 1a.

In the following, we present our experimental data obtained for this material. Figure 1b shows the optical image of ~ 7 nm thick MoS$_2$ with four-probes on the SiO$_2$ (300 nm) / Si (heavily p-doped) substrate. Figure 1c displays the resistivity $\rho$ in the unit of quantum resistance $h/e^2$ for the chosen back-gate biases, extracted in the zero drain-source voltage limits. Observing two sharply distinct families of curves $\rho(T)$, $\Delta\rho/\Delta T > 0$ and $\Delta\rho/\Delta T < 0$ suggests the metal-insulator transition at a critical backgate bias $V_c \sim 21$ V at low temperature. We also note the pronouncedly non-monotonic temperature dependence of resistivity in the metallic phase. The peak value near the transition exceeds the upper bound $\rho_c \lesssim (2/n_s n_v)h/e^2$ predicted by the Mott-Ioffe-Regel theory[22], where $n_s$ is the spin and $n_v$ the valley degeneracy: $n_s = 2$ and $n_v = 6$ for multilayer MoS$_2$[23]. Also, the resistivity maximum occurs at the temperature $T^* \sim 20$ K much lower than the Fermi temperature $T_F \sim 200$ K. Very similar behavior of the resistivity has been reported in organic Mott systems[24] and only in particularly clean 2D systems[1,25], and it has been interpreted to originate from strong correlation effects. In some sense, the observation of such features in multilayer MoS$_2$ of rather low field effect mobility $\mu \approx 1500$ cm$^2$V$^{-1}$s$^{-1}$ at 2 K (see Supplementary Fig. 1 for the temperature dependent field effect mobility) might, at a first glance, look surprising. In addition, the Fermi energy in this system is $T_F \sim 200$ K, much higher compared to standard high mobility 2DEG where it is typically around 10 K. From the theoretical perspective, however, only relative energy scales are significant, and those are comparable in both systems. As we mentioned earlier, MoS$_2$ has a strong correlation due to the large effective mass and poor screening.

In order to check the possibility of Mott quantum criticality in this material, we next perform the scaling analysis of the resistivity maxima based on the theoretical predictions for the Hubbard model. Assuming the resistivity of the form, $\rho(T) = \rho_0 + \delta\rho(T)$ where $\rho_0$ is the residual resistivity due to impurity scattering and $\delta\rho(T)$ is the temperature dependent resistivity dominated by electron-electron scattering, the scaling form was proposed as[14],

$$\delta\rho(T) = \delta\rho_{max} f(T/T_{max}), \qquad (1)$$

where $\delta\rho_{max} = \rho_{max} - \rho_0$ and $\rho_{max}$ is the maximum resistivity at $T = T_{max}$.

Following this scheme, we now perform the scaling of resistivity for five different backgate biases $V_{bg}$, as shown in Fig. 1d, and they display an excellent collapse. When we compare the collapsed data to the scaling curve theoretically predicted for a single band Hubbard model at half filling (blue curve, which also coincides with collapsed curve of 2DEG in Silicon)[14], we do notice a certain systematic discrepancy. In the following, we argue that this discrepancy is due to disorder that is significantly higher than in 2DEG. For this purpose, we extended previous theoretical calculation to include moderate disorder, and investigate how this affects the scaling of the resistivity maxima. We consider a disordered Hubbard model at half filling, where disorder is modeled as a site diagonal random potential chosen from uniform distribution in the energy range [-$W$/2, $W$/2]. We treat disorder on the level of coherent potential approximation (CPA)[26], treating the correlation effects using DMFT theory, the same technique as in previous work (see Methods). Moderate strength of disorder of $W = 2.5$, in the unit of half-bandwidth, allows us to match disordered DMFT scaling function with our collapsed data. Now we clearly see the effects of disorder, i.e., it simply produces a slight broadening of the scaling function, reflecting the effect of disorder to broaden the bandwidth of the electronic spectrum[27]. These disorder effects, however, are only quantitative changes that should not introduce any significant qualitative modifications to the Quantum Critical Behavior at the Mott point[27]. The use of CPA approximation here is completely justified, since the considered disorder level is moderate and we should be far from Anderson (disorder-driven) metal-insulator transition. In addition, the observed power law dependence of $T_{\max}$ as a function of the reduced carrier density $\delta n = (n - n_c)/n_c$ as shown in Fig. 1e supports this expectation. This remarkable scaling description strongly suggests the Mott-like transition in multilayer $MoS_2$.

From a different point of view, Finkel'stein and Punnoose[28,29] developed a theory that describes the non-monotonic temperature dependence of resistivity as a result of the competition between disorder and interactions. In this theory, the metallic state is stabilized by the strong correlation and the localization is driven by disorder. However, we find that the scaling based on this theory is poor (see Supplementary Information), making us more confident to pursue more analysis for Mott criticality in this multilayer $MoS_2$.

Previous scaling analyses of the experimental results on 2D MIT have been made along the temperature-independent critical transition-driving parameter, i.e., $n$ or $V_{bg}$ in our case, as schematically shown in Fig. 1f (red dashed line). The Mott transition is, however, a first-order phase transition at low temperature, and the critical scaling behaviour is expected above the critical end point $T_c$. In this case, for the quantum critical scaling, it is essential to identify the instability trajectory where the system is least stable, called the quantum Widom line (QWL)[16,30,31], which separates metal and insulator, and is usually temperature dependent as shown in Fig. 1f (red solid line). There are several ways to define the QWL but it is known that the scaling behaviour is not so sensitive to the choice of them[31]. We find the QWL from the inflection points in $\log\rho(V_{bg}, T)$ vs. $V_{bg}$ curves. To identify the inflection points, $(V_c(T), \rho_c(T))$, we firstly fit the data $\log\rho(V_{bg}, T)$ by a polynomial for each temperature, and differentiate it with respect to $V_{bg}$ to find the maximum (Supplementary Information). Figure 2 shows the consequential phase diagram. The dashed line is the interpolation to the $V_c \sim 21$ V at the lowest temperature. The QWL exhibits the same non-monotonic form as in organic systems[15]. In these organic systems in which the phase transition is driven by the pressure, the positive slope of the QWL reflects the large entropy of localized spins on the insulating side, consistent with Clausius-Clapeyron equation. The close similarity in Fig. 2 strongly suggests the formation of local moment, i.e., Mott localization in this multilayer $MoS_2$. The slope change at high temperature is due to the thermal filling of the Mott gap. Two blue points define the boundary between the bad metal and the Fermi liquid obtained from the inflection points in $\rho(V_{bg}, T)$ vs. $T$ (Supplementary information)[31].

Based on the QWL in Fig. 2, we plot the normalized resistivity $\rho/\rho_c$ as a function of $\delta V_{bg}$ for several temperatures in Fig. 3a, where $\delta V_{bg} = V_{bg} - V_c(T)$. All traces cross at $\delta V_{bg} = 0$, transition point between metal and insulator. Then, we rescale $\delta V_{bg}$ for each trace at different temperature by $w(T)$ so that all curves collapse into a universal one in $\rho/\rho_c$ vs. $\delta V_{bg} w(T)$ as shown in Fig. 3b. The inset in this figure displays the scaling parameter $w$ for temperature in log scale. The dynamical mean-field theory (DMFT) predicts that the resistivity scaling holds above $\sim 2T_c$. In fact, the linear, i.e., power law behaviour in the inset appears in the temperature range, $20 \lesssim T \lesssim 80$ K. We extract the exponent $z\nu$ from the slope via $w(T) \sim T^{-1/z\nu}$, where $\nu$ is the critical exponent of the correlation length and z is the dynamical critical exponent. From the linear fit, we obtain $z\nu = 2.56$.

As usually expected for quantum criticality, the resistivity should follow the scaling relation, $\rho(T, \delta V) = \rho_c(T) f(T/T_0(\delta V))$, where $T_0 = |c\delta V|^{z\nu}$ with an arbitrary constant $c$. With a value $z\nu = 2.56$ along the QWL, all curves of $\rho/\rho_c$ in the scaling regime of temperature collapse into two branches of insulator and metal, respectively, as shown in Fig. 4. The scaling along the temperature-independent $V_c = 21\,\text{V}$ causes non-equal $z\nu$ values for the insulating and the metallic sides (see Supplementary Information). We believe that this is the result from the incorrect scaling for the Mott criticality in this system. The scalings for all reported 2D MIT phenomena have been performed in the latter scheme. The asymmetric $z\nu$ across the 2D MIT in a certain system might be amended in a Mott scaling picture[32].

The critical exponent $z\nu = 2.56$ for multilayer $MoS_2$ is rather large, different than in other systems in which the Mott transition is speculated. The DMFT calculation estimates[16] $z\nu = 0.56$~$0.57$, and the experiments in organic materials report the similar values[15], $z\nu = 0.49$~$0.68$. Although the organic materials and our multilayer $MoS_2$ share the several similar Mott-features, the distinctly different $z\nu$ suggests that they belong to very different universality classes. In fact, the DMFT predicts $z\nu = 1.35$ for the doped case[30], in which the transition is driven by electrically doped the system, as usually have been implemented in 2D systems such as Si-MOSFETs. This theory is based on the single band Hubbard model without inter-site interactions. The values of $z\nu$ in clean Si-MOSFETs[1,2] are 1.2~1.6 comparable to the above theoretical value, but we note that these experimental results were not based on the Mott picture using Widom line.

**Conclusion**

Including inter-site interactions, the Wigner crystallization similar to Mott physics can occur for any partially filled band in the low carrier-density limit, requiring charge ordering with local magnetic moments[13]. The precise role of such charge ordering in the vicinity of the transition has not been investigated from the theoretical perspective although most reported experimental results in 2D systems correspond to this type of transition. Besides, the difference in the underlying band structure or the form of the lattice is known to likely originate various Mott systems with different critical exponents.

The Mott transition in 2D systems has not been clearly demonstrated. Even in the cleanest Si-MOSFETs, the non-monotonic temperature dependent resistivity in the metallic phase near the MIT has often been interpreted as a result of a dominating correlation effects in disorder-driven localization. In this disorder scenario, the scaling behaviour by the normalization group equation generally applies to only few traces near the MIT[8,29]. We showed that the non-monotonic behaviour in multilayer $MoS_2$ is excellently explained by the scaling in the Mott picture with disorder for the wide range of carrier density but poorly in disorder scenario. Moreover, the successful scaling with one $z\nu$ value for both metallic and insulating phases using the temperature dependent Widom line strongly suggests the Mott-like transition in a 2D system.

**Methods**

**Device fabrication and electrical transport measurement.** A multilayer $MoS_2$ flake was mechanically exfoliated on a highly p-doped 300 nm $SiO_2$ /Si substrate. Four electrodes were patterned using electron beam lithography (EBL) after PMMA A4 was spin-coated (3000 rpm, 50 s). Then, Cr/Au (1/60 nm) were evaporated in high vacuum ($\sim 10^{-6}$ Torr). The dimension of the sample (length = 5.5 μm, width = 4.5 μm, and thickness = $\sim$ 7 nm ) was confirmed by an atomic force microscope (SPA 400, SEIKO). Four-probe electrical measurements were performed using a commercial characterization system (B1500A, Keysight Technologies) in a cryostat (PPMS, Quantum Design, Inc.).

**Resistivity scaling curve of disordered Hubbard model.** For the purpose of investigation of disorder effects in strongly correlated systems we consider the half-filled single-orbital Hubbard model with site-diagonal disorder and nearest neighbour hopping.

$$\mathrm{H} = -\sum_{i,j,\sigma} t_{i,j} c_{i\sigma}^\dagger c_{j\sigma} + U \sum_i n_{i\uparrow} n_{i\downarrow} + \sum_{i\sigma} v_i n_{i\sigma} - \mu \sum_{i\sigma} n_{i\sigma}, \qquad (2)$$

where $t_{i,j}$ is nearest neighbour hopping amplitude, $c_{i\sigma}^\dagger$ ($c_{i\sigma}$) are charge carriers creation (annihilation) operators, $n_{i\sigma} = c_{i\sigma}^\dagger c_{i\sigma}$ is the occupation number operator on site *i* for spin *σ*. Interaction is described by interaction strength parameter $U$, $\mu$ is chemical potential chosen to impose half filling, and $v_i$ is the random on-site potential (source of disorder). We choose $v_i$ from uniform distribution, since most of the features of the disordered Hubbard model are expected to be insensitive to the particular form of the disorder distribution.

In the spirit of DMFT, DHM can be reduced to the ensemble of Anderson impurity models in a self-consistently determined conduction bath. Here we consider an ensemble of impurities, and we use coherent potential approximation (CPA) to close the set of self-consistent equations of disordered Hubbard model. Within the CPA we solve an impurity model for each on-site energy, and we average Green's functions

$$G_{av}(i\omega) = \frac{1}{N}\sum_{i=1}^{N} G_i(i\omega). \qquad (3)$$

From averaged Green's function we calculate dynamical mean field and from this point DFMT equations are analogous to those in the clean case.

Optical conductivity $\sigma$ is calculated using Kubo formula, and dc resistivity is defined as

$$\rho_{dc} = [Re\ \sigma(\omega = 0)]^{-1} \qquad (4)$$

Further, we apply scaling ansatz (we divide resistivity by maximum of the resistivity $\rho/\rho_{max}$ and temperature by temperature where the maximum is reached $T/T_{max}$) to all metallic curves close to the transition, and we collapse all resistivity curves to the single scaling curve of DHM. The scaling curve is well defined outside of Fermi-liquid region.

24. Limelette, P. *et al.* Mott Transition and Transport Crossovers in the Organic Compound κ−(BEDT−TTF)$_2$Cu[N(CN)$_2$]Cl. *Phys. Rev. Lett.* **91,** 016401 (2003).

25. Hanein, Y., Meirav, U., Shahar, D., Li, C., Tsui, D. & Shtrikman, H. The metalliclike conductivity of a two-dimensional hole system. *Phys. Rev. Lett.* **80,** 1288-1291 (1998).

26. Dobrosavljević, V. & Kotliar, G. Strong correlations and disorder in d=∞ and beyond. *Phys. Rev. B* **50,** 1430-1449 (1994).

27. Radonjić, M. M., Tanasković, D., Dobrosavljević, V. & Haule, K. Influence of disorder on incoherent transport near the Mott transition. *Phys. Rev. B* **81,** 075118 (2010).

28. Finkelshtein, A. Influence of Coulomb interaction on the properties of disordered metals. *Sov. Phys. JETP* **57,** 97-108 (1983).

29. Punnoose, A. & Finkel'Stein, A. M. Dilute electron gas near the metal-insulator transition: Role of valleys in silicon inversion layers. *Phys. Rev. Lett.* **88,** 016802 (2001).

30. Vučičević, J., Tanasković, D., Rozenberg, M. & Dobrosavljević, V. Bad-metal behavior reveals Mott quantum criticality in doped Hubbard models. *Phys. Rev. Lett.* **114,** 246402 (2015).

31. Vučičević, J., Terletska, H., Tanasković, D. & Dobrosavljević, V. Finite-temperature crossover and the quantum Widom line near the Mott transition. *Phys. Rev. B*, **88,** 075143 (2013).

32. Moon, B. H. *et al.* Soft Coulomb gap and asymmetric scaling towards metal-insulator quantum criticality in multilayer MoS$_2$. *Nat. Commun.* **9,** 2052 (2018).

**Acknowledgments**

This work was primarily supported by the Institute for Basic Science (IBS-R011-D1). V.D. acknowledges that work in Florida was supported by the NSF Grant No. 1822258, and the National High Magnetic Field Laboratory through the NSF Cooperative Agreement No. 1157490 and the State of Florida. M.M.R. acknowledges support from the Ministry of

**Additional information**

Supplementary information is available in the online version of the paper. Reprints and permissions information is available online at www.natrue.com/reprints.

Correspondence and requests for materials should be addressed to B.H.M. or V.D.

**Competing financial interests**

The authors declare no competing financial interests.

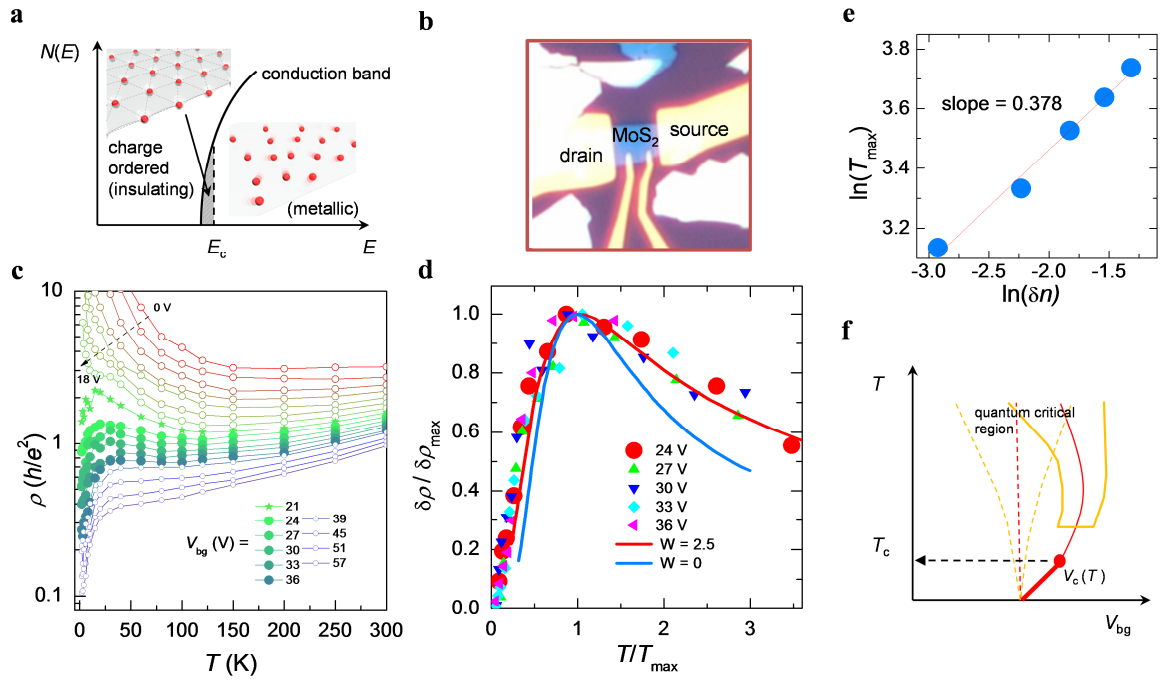

**Figure 1 | Temperature dependent resistivity and scaling plot. a,** Schematic view of metal-insulator transition. **b,** Optical image of a multilayer $MoS_2$. **c,** Electric resistivity ($\rho$) as a function of temperature ($T$) for various backgate biases $V_{bg}$. **d,** Normalized resistivity as a function of normalized temperature for five consecutive $V_{bg}$. The blue and the red solid lines are the theoretical scaling results from the Hubbard model with a disorder strength $W = 0$ and 2.5, respectively. **e,** $T_{max}$ as a function of a reduced carrier density. **f,** Schematic phase diagrams and quantum critical regions for the conventional MIT case (temperature independent $V_c$) and the Mott MIT (temperature dependent $V_c$).

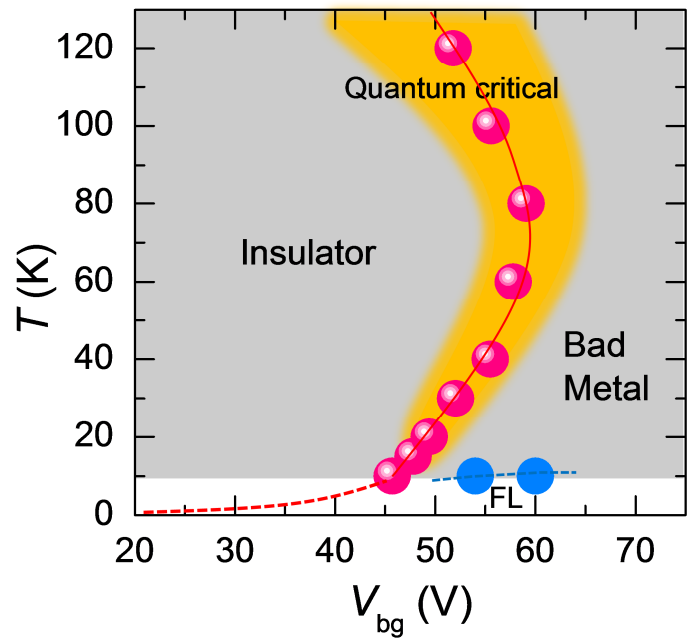

**Figure 2 | Phase diagram.** Phase diagram of multilayer $MoS_2$. The red dashed line is the simple interpolation to $V_c \sim 21$ V at the lowest temperature.

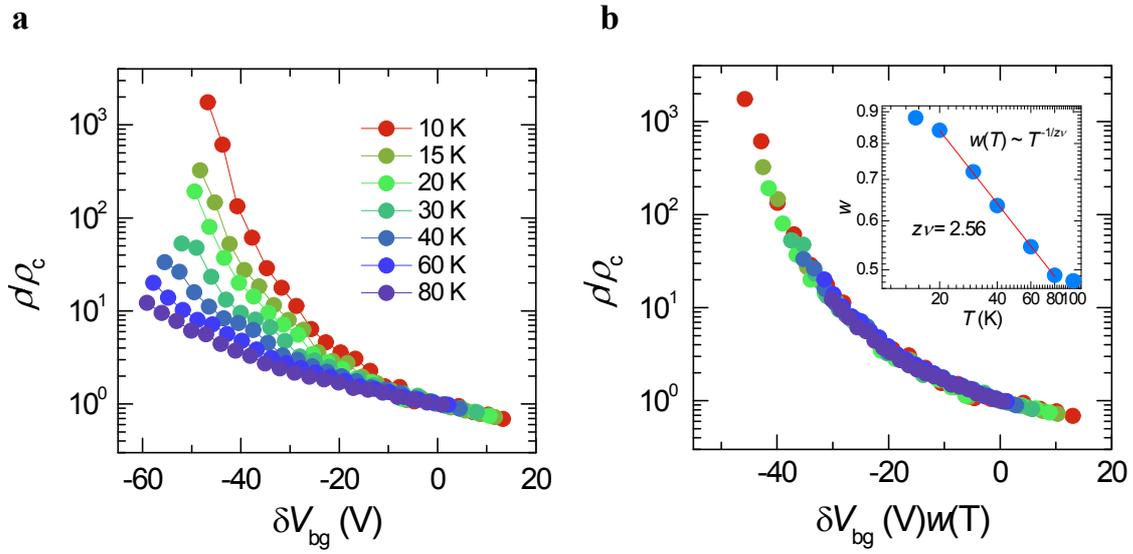

**Figure 3 | Scaling using scaling parameter. a,** The normalized resistivity for the adjusted backgate bias $\delta V_{bg} = V_{bg} - V_c(T)$ by the critical backgate bias $V_c(T)$ along the QWL. **b,** Scaling plot of normalized resistivity for rescaled $\delta V_{bg}$ by scaling parameter $w(T)$. The inset show the scaling parameter as a function of temperature in log-log scale.

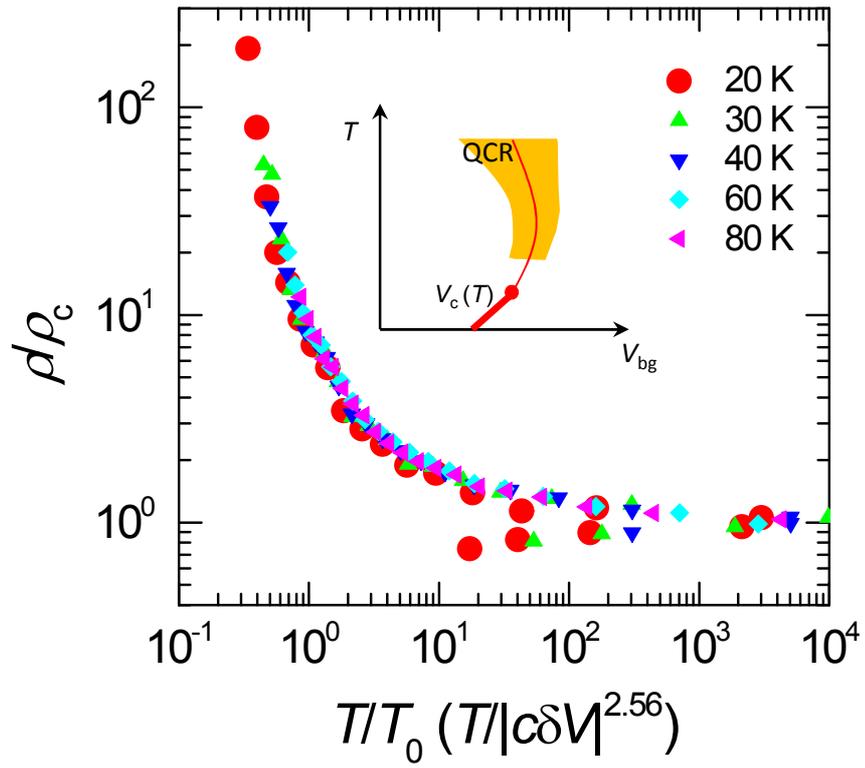

**Figure 4 | Scaling plot.** Scaling of normalized resistivity for normalized temperature by $T_0 = |c\delta V|^{z\nu}$ with the given value of $z\nu = 2.56$.